\documentclass[12pt]{article}
\usepackage{sectsty}
\usepackage{mathrsfs}
\allsectionsfont{\sffamily}
\newcommand{\x}{arXiv:}

\usepackage[nosort]{cite}

\usepackage[pdftex]{graphicx}
\usepackage[font={scriptsize,singlespacing}]{caption}
\usepackage{epstopdf}

\usepackage{amsmath}
\usepackage{amssymb}
\usepackage{subfigure}
\usepackage{hyperref}
\usepackage{url}
\usepackage{xcolor}
\usepackage{color}
\definecolor{amaranth}{rgb}{0.9, 0.17, 0.31}
\definecolor{purple(munsell)}{rgb}{0.62, 0.0, 0.77}
\definecolor{americanrose}{rgb}{1.0, 0.01, 0.24}
\definecolor{palatinateblue}{rgb}{0.15, 0.23, 0.89}
\definecolor{royalblue(web)}{rgb}{0.25, 0.41, 0.88}
\definecolor{hanpurple}{rgb}{0.32, 0.09, 0.98}
\definecolor{beaublue}{rgb}{0.74, 0.83, 0.9}
\definecolor{carminered}{rgb}{1.0, 0.0, 0.22}
\definecolor{brightpink}{rgb}{1.0, 0.0, 0.5}
\hypersetup{ linktoc=all,
    colorlinks, linkcolor={palatinateblue},
    citecolor={brightpink}, urlcolor={purple(munsell)}
}

\def\sideremark#1{\ifvmode\leavevmode\fi\vadjust{\vbox to0pt{\vss
 \hbox to 0pt{\hskip\hsize\hskip1em
 \vbox{\hsize2cm\tiny\raggedright\pretolerance10000
 \noindent #1\hfill}\hss}\vbox to8pt{\vfil}\vss}}}%
\newcommand{\bo}{\raise-1mm\hbox{\Large$\Box$}}

\newcommand{\be}{\begin{equation}}
\newcommand{\ee}{\end{equation}}
\newcommand{\bea}{\begin{eqnarray}}
\newcommand{\eea}{\end{eqnarray}}

\begin{document}
\thispagestyle{empty}
\begin{center}

\null \vskip-1truecm \vskip2truecm

{\Large{\bf \textsf{Hawking Evaporation Time Scale
of \vskip0.2truecm Topological Black Holes in Anti-de Sitter Spacetime}}}

\vskip1truecm

{\textsf{Yen Chin Ong}}\\
\vskip0.1truecm
{Nordita, KTH Royal Institute of Technology and Stockholm University, \\ Roslagstullsbacken 23,
SE-106 91 Stockholm, Sweden}\\
{\tt Email: yenchin.ong@nordita.org}\\

\end{center}
\vskip1truecm \centerline{\textsf{ABSTRACT}} \baselineskip=15pt

\medskip

It was recently pointed out that if an absorbing boundary condition is imposed at infinity, an asymptotically anti-de Sitter Schwarzschild black hole with a spherical horizon takes only a finite amount of  time to evaporate away even if its initial mass is arbitrarily large. We show that this is a rather generic property in AdS spacetimes: regardless of their horizon topologies, neutral AdS black holes in general relativity take about the same amount of time to evaporate down to the same size of order $L$, the AdS length scale. Our discussion focuses on the case in which the black hole has toral event horizon. A brief comment is made on the 
hyperbolic case, i.e. for black holes with negatively curved horizons.

\section{Finite Upper Bound for Hawking Radiation Time}\label{S1}

Anti-de Sitter (AdS) spacetime plays an important role in theoretical physics \cite{gibbons}, especially in the holographic duality between AdS spacetime and conformal field theory (CFT) \cite{mal}. 
As is well known, AdS spacetime is not globally hyperbolic, and 
one needs to impose some boundary conditions at infinity. If the usual reflective boundary condition is chosen, a light ray from an arbitrary ``center'' in the bulk can reach the boundary and be reflected back in a finite proper time of an observer sitting at said ``center''. 
A large\footnote{``Large'' means the size of the black hole is larger than the AdS length scale $L$.} black hole in the bulk therefore tends not to evaporate, but instead achieve thermal equilibrium with its own Hawking radiation that gets reflected back from infinity. 

However, one could choose an absorbing boundary condition instead, say by coupling
the boundary field theory with an auxiliary system (``AUX''), such as another CFT. 
(In quantum field theory, boundary conditions are also required for quantization  in a non-globally hyperbolic manifold. See  \cite{AIS} for a discussion of ``transparent'' vs. ``reflective'' boundary conditions and the various quantization schemes in AdS spacetime. The boundary condition also affects whether a given asymptotically AdS spacetime is stable under small perturbation \cite{1312.5544,1104.3702}.)
With such a ``CFT-AUX'' system at work,
even large AdS black holes can evaporate \cite{1304.6483,0804.0055,1307.1796,1403.4886}. Dynamical and non-equilibrium scenarios are of great interest in holography \cite{chesler}, especially in the applications to material systems like condensed matter and quark gluon plasma. The understanding of the behaviors of evaporating large black holes is a crucial step toward this goal.

\begin{figure}[!h]
\centering
\mbox{\subfigure{\includegraphics[width=3.2in]{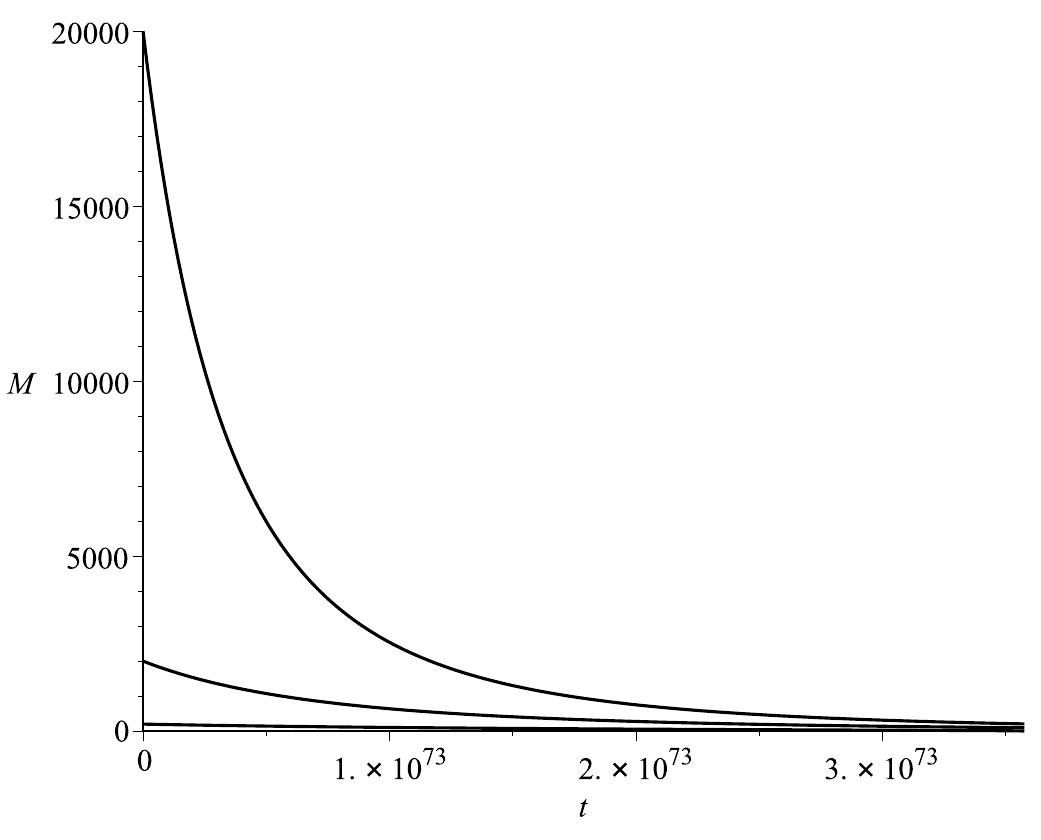}}\quad
\subfigure{\includegraphics[width=3.2in]{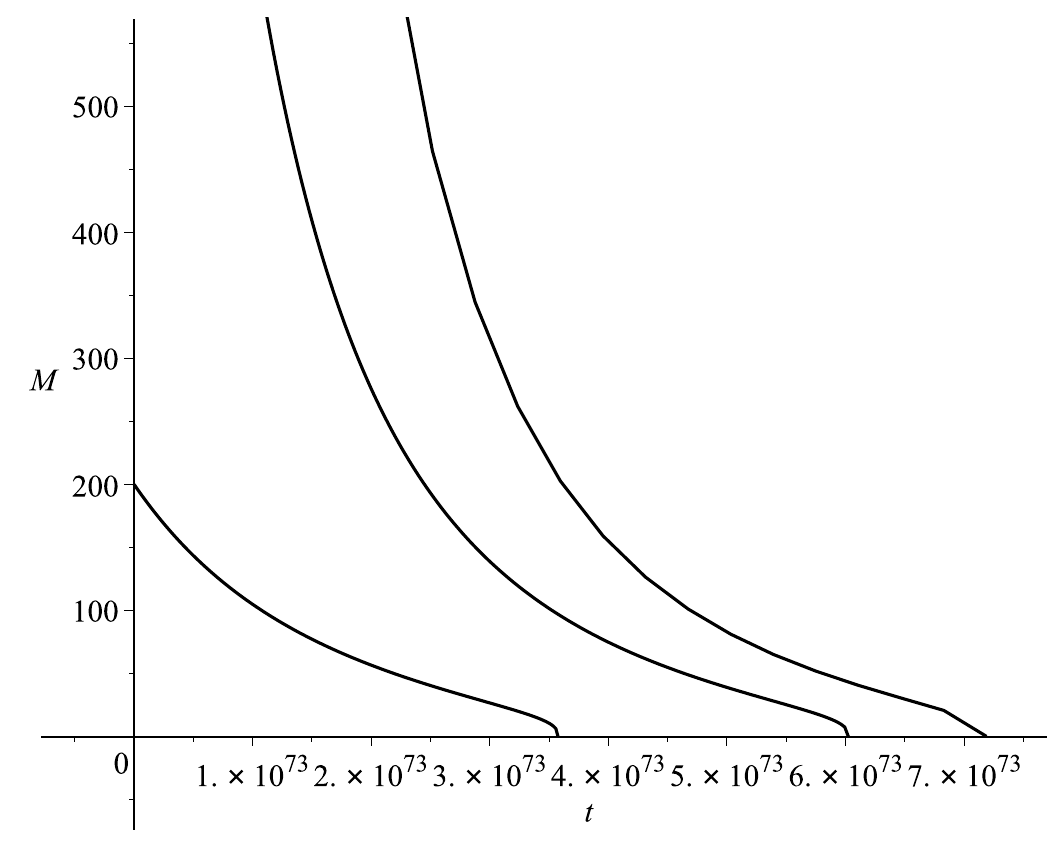} }}
\caption{\textbf{Left:} The evolution of some AdS-Schwarzschild black holes with an absorbing boundary condition imposed at infinity. In this example, we set the numerical value $L=100$, and the initial masses of the black holes are 200, 2000, and 20000, respectively. \textbf{Right:} A closer look toward the end of the evaporation, from which we see that these black holes reach the zero mass limit at about the same time of order $L^3/\hbar$, within an order of magnitude or so. The evolution should only be trusted quantitatively upto $M \sim L$ beyond which the geometric optics approximation is no longer valid.
The following comments are true also for the other figures that show mass evolution of the various black holes in this work: since we have neglected the greybody factors, the lifetime is expected to be off by a few magnitudes anyway. Note also that in the units in which length is in centimeters, and $G=1=c$, $\hbar = \hbar G/c^3 \approx  3 \times  10^{-66} \text{cm}^2$. 
\label{k1}}
\end{figure}

In a recent work by Don Page \cite{1507.02682}, it was shown that an asymptotically anti-de Sitter black hole with a standard spherical horizon of $S^2$ topology equipped with the canonical round metric (hereinafter, ``AdS-Schwarzschild black hole'') takes a time proportional to $L^{3}$ to evaporate away. Some numerical examples are provided in Fig.(\ref{k1}). These plots assume the mass loss of the black holes follow the geometric optics approximation, which of course is only true for large mass regime $M \gg L$. In other words, the evolution of the masses beyond $M \sim L$ should not be trusted quantitatively in the plots, though it is still \emph{qualitatively} correct. As explained in \cite{1507.02682}, the evolution from $M \sim L$ down to $M=0$ should take a time of around $t \sim L$).

More specifically, in $d$-dimensions, the evaporation time scale is \cite{1507.02682}
\begin{equation}\label{eq1}
t_{\text{evap}} = \mathcal{C}\frac{L^{d-1}}{\hbar},
\end{equation}
where $\mathcal{C}$ depends on the spacetime dimension and on the field content of the theory.
In this work, we will work with the units such that $c=G=1$ but $\hbar \neq 1$. This differs from the convention in \cite{1507.02682}.

This result implies that even an \emph{arbitrarily large} AdS-Schwarzschild black hole takes only a finite time, \emph{fixed by the cosmological constant}, to evaporate away. This should be contrast with the case of an asymptotically flat Schwarzschild black hole with an initial mass $M$, whose evaporation time is proportional to $M^3$, and therefore an arbitrarily large hole takes an arbitrarily long time to evaporate away.

This is not the first time we see that in some instances, it is the cosmological length scale $L$, instead of the black hole mass, that characterizes some of the physical properties of topological black holes.
However, as we will soon discuss, most of the previous observations \cite{1403.4886} involved AdS black holes with flat event horizons, i.e. black holes that are the most important in applied holography.
In fact, it is well known that AdS black holes can have horizons which are positively curved, flat, and also negatively curved \cite{9709039, 9808032, 9705004, 0011097}. For earlier related works, see also, \cite{9404041,9407024,9604005}.
The metric tensor for an asymptotically locally (neutral) AdS black hole in $d=n+2$ dimensions takes the form 
\begin{equation}\label{metric}
g[\text{AdS}(k)]=-\left(k + \frac{r^2}{L^2} - \frac{16\pi M}{n V[X^k_n]r^{n-1}}\right)dt^2 + \left(k + \frac{r^2}{L^2} - \frac{16\pi M}{n V[X^k_n]r^{n-1}}\right)^{-1} dr^2 + r^2d\Omega^2[X^k_n],
\end{equation}
where $L$ is the AdS length scale, and $d\Omega^2[X^k_n]$ is a Riemannian metric of constant curvature $k=\left\{-1,0,+1\right\}$ on the orientable manifold $X^k_n$, and $V[X^k_n]$ is the dimensionless area of this space. For example, for $k=1$ and $n=2$, the underlying space is $X^k_n= S^2$ and the dimensionless area is $4\pi$. The space $X^k_n$ is compact unless otherwise specified.

The Hawking temperature of these black holes are given by the general formula \cite{9808032, 0011097},
\begin{equation}\label{temp}
T = \frac{\hbar}{4\pi L^2 r_h}\left[(d-1)r_h^2 + (d-3)kL^2\right],
\end{equation}
where $r_h$ denotes the horizon.
Although the main focus of this work is on the $k=0$ case, 
we will also make some comments on the negatively curved case towards the end. 

Let us mention a few previous observations that AdS length scale $L$, instead of the black hole mass, has characterized some properties of the toral black holes \cite{1403.4886}. In these examples, the spacetime is 4-dimensional. 
\begin{itemize}
\item[(1)] The maximal in-falling time $\tau_{\text{max}}$ from the horizon to the singularity for a neutral toral black hole is fixed by $L$, not the black hole mass $M$:
\begin{equation}\label{wow}
\tau_{\text{max}} = \int_0^{r_h} \left(\frac{2M}{\pi K^2r} -\frac{r^2}{L^2}\right)^{-\frac{1}{2}} dr = \frac{\pi L}{3}, ~~r_h = \left(\frac{2ML^2}{\pi K^2}\right)^{\frac{1}{3}}.
\end{equation}
(Here $K$ is a compactification parameter of a torus, see metric \ref{toral} below.)\newline
In contrast, an asymptotically flat Schwarzschild black hole has maximal in-falling time given by
\begin{equation} \tau_{\text{max}}= \int_0^{r_h} \left(\frac{2M}{r} -1 \right)^{-\frac{1}{2}}dr = \pi M; ~~r_h=2M. \notag \end{equation}
\item[(2)] The Kretschmann scalar $R^{\mu\nu\alpha\beta}R_{\mu\nu\alpha\beta}$ at the event horizon for a neutral toral black hole is $36/L^4$, which is also independent of $M$. (The extremally charged toral case gives $144/L^4$ at the horizon.)
For an asymptotically flat Schwarzschild black hole, on the other hand, the Kretschmann scalar at the event horizon is $0.75/M^4$.
\end{itemize}

It is therefore interesting to ask if the observation in \cite{1507.02682} also generalizes to black holes with toral event horizons. In this work, we find that it does, modulo some subtle differences.

\section{A Subtler Case for Toral Black Holes}

Let us consider a 4-dimensional neutral black hole with toral horizon. Its metric tensor is 
\begin{equation}\label{toral}
g[\Bbb{T}^2\text{-AdS}]=-\left(\frac{r^2}{L^2}-\frac{2M}{\pi K^2 r}\right)dt^2 + \left(\frac{r^2}{L^2}-\frac{2M}{\pi K^2 r}\right)^{-1}dr^2 + r^2(d\zeta^2 + d\xi^2),
\end{equation}
where $\zeta, \xi \in [0,2\pi K)$ are coordinates on a flat square torus $\Bbb{T}^2=\Bbb{R}^2/\Bbb{Z}^2$. The horizon therefore has an area $4\pi^2K^2 r_h^2$, and $K$ plays the role of a ``compactification parameter''. In general the torus can be of other shapes (not necessarily a square), and in fact in higher dimensions, some quotients of tori (e.g. $\Bbb{T}^2/\Bbb{Z}_2$) are permitted, but for simplicity we shall focus on 4-dimensional flat square tori.

The Hawking temperature for this toral black hole is \cite{1403.4886}
\begin{equation}
T=\frac{3\hbar r_h}{4\pi L^2}=\frac{3\hbar M}{2\pi^2 K^2 r_h^2}; ~~r_h = \left(\frac{2ML^2}{\pi K^2}\right)^{\frac{1}{3}}.
\end{equation}

For an asymptotically flat Schwarzschild black hole (and also for a Schwarzschild-AdS black hole), the effective potential for massless particle has a local maximum at the photon orbit $r=3M$, and one uses this photon sphere as the emitting surface in the geometric optic approximation. Of course, as emphasized in \cite{1507.02682}, this does not give a precise lifetime; to do this one has to explicitly compute the greybody factors\footnote{For a study of greybody factors for AdS black holes, see e.g., \cite{0708.0017}. However their analysis assumes a reflective boundary condition.} of the various particles in the theory. In this work, following \cite{1507.02682}, we shall also ignore the greybody factors, since we are only concerned with the qualitative features of the evaporation.

As discussed in \cite{1403.4886,9803061}, the effective potential for massless particles in the background of a toral black hole geometry does \emph{not} have a local maximum. For an emitted particle with angular momentum $J$, the potential is a monotonically increasing function of the coordinate radius $r$:
\begin{equation}
V(r) = \frac{J^2}{r^2}\left[\frac{r^2}{L^2} - \frac{2M}{\pi K^2 r}\right].
\end{equation}
This function tends to a constant value $J^2/L^2$ as one gets close to the boundary $r=\infty$. It turns out that the relevant area $A$ that one should use in the Stefan-Boltzmann law $L \propto AT^4$  is in fact $4\pi^2K^2 L^2$, which is fixed by the cosmological constant instead of the black hole mass \cite{1403.4886, 9803061}.

The mass loss equation is then (see also \cite{1503.08245}), up to greybody factors, and in the geometric optics regime,
\begin{equation}\label{ODE}
\frac{dM}{dt} = -a\pi^2K^2L^2 \left[\frac{3\hbar M}{2\pi^2K^2 r_h^2}\right]^4 = -\mathcal{B}M^\frac{4}{3},
\end{equation}
where $a=\pi^2/15\hbar^3$ is the radiation constant.
We have separated $M$ from all the other factors, which we have simply denoted by $\mathcal{B}$. It is, explicitly,
\begin{equation}\label{B}
\mathcal{B}=\frac{27\hbar}{2^{\frac{20}{3}}5 \pi^{\frac{4}{3}}} K^{-\frac{2}{3}} L^{-\frac{10}{3}}.
\end{equation}
For later convenience, let us define a dimensionless quantity $\mathcal{C}$ by
\begin{equation}\label{C}
\mathcal{B} = \frac{\hbar}{\mathcal{C}} L^{-\frac{10}{3}}.
\end{equation}
Solving the differential equation \ref{ODE}, one arrives at 
\begin{equation}\label{sol}
M(t) = \left[\frac{3}{\mathcal{B}t+3M_0^{-\frac{1}{3}}}\right]^3,
\end{equation}
where $M_0 \equiv M(t_0 = 0)$ is the initial mass.

One immediately sees that unlike the AdS-Schwarzschild black hole, toral black holes only tend to zero mass asymptotically\footnote{This statement ignores other effects that could affect the black hole geometry, such as the phase transition to Horowitz-Myers soliton \cite{9808079,0101134,0108170,0205001} for a sufficiently cold toral black hole. For some applications of this phase transition, see \cite{1403.4886,0910.4456}. There is no phase transition to AdS background \cite{9808032,9705004,9809029}.}. Naively then, any two such black holes with different initial masses would eventually get very close to zero mass if one waits long enough. Note that, however, this is assuming the mass loss equation as given by the geometric optics approximation continues to hold. In view of these,
in order to compare with the result in \cite{1507.02682} for the AdS-Schwarzschild case, we have to phrase the question slightly differently. Note that in \cite{1507.02682}, it was emphasized that the geometric optic approximation is good until around $r_h \sim L$, and the black holes take a time proportional to $L^3$ to shrink down to \emph{that} size. We could then ask the same question for the toral case: 
\begin{itemize}
\item
For a fixed compactification parameter (say $K=1$), how long does it take for a toral black hole of a given mass $M > L$ to shrink down to $M=L$?  
\end{itemize}
Indeed, as found in \cite{9803061}, the geometric optics approximation is good for the toral case also for $M \gg L$.

As we will see below, the result is indeed similar: toral black holes with initial mass greater than $L$ take a time proportional to $L^3$ to reach the mass $M=L$. Thus, in a very loose sense, they ``converge'' to the line $M=L$ at about the same time (this is of course an order of magnitude estimate; $2 \times 10^{66}$, for example, is ``close'' to $5 \times 10^{66}$ in this sense). 

\begin{figure}[!h]
\centering
\mbox{\subfigure{\includegraphics[width=3.3in]{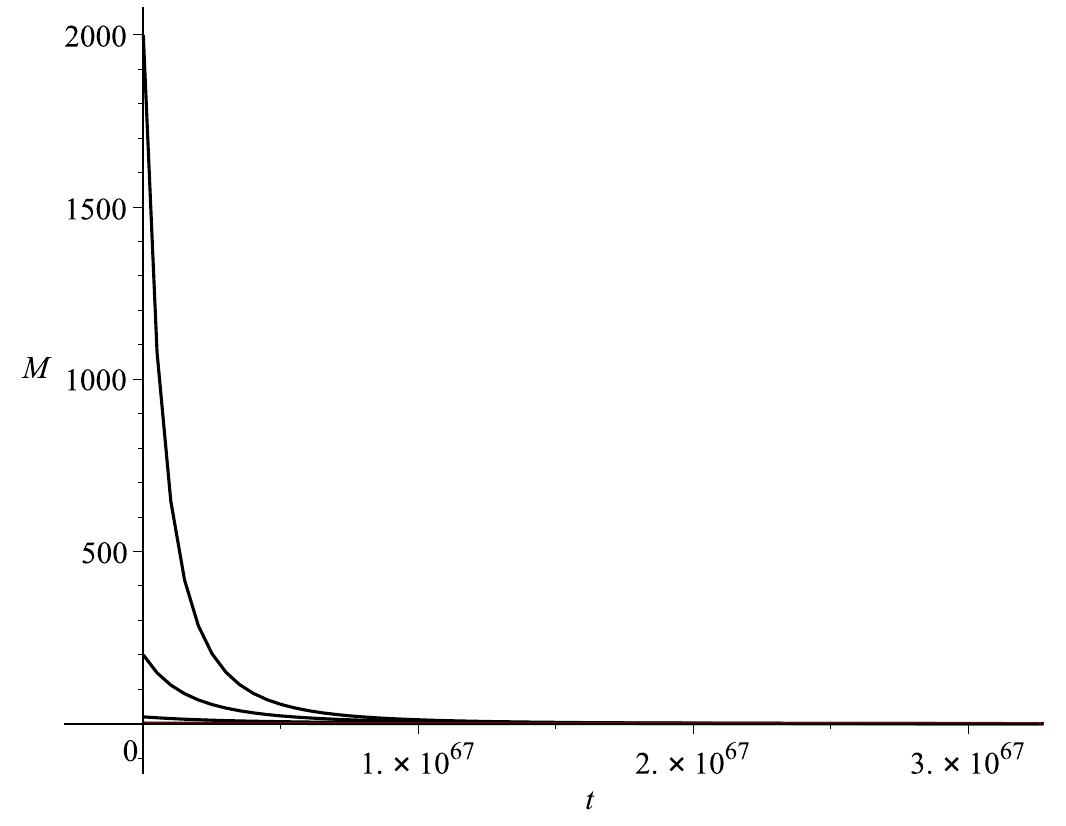}}\quad
\subfigure{\includegraphics[width=3.1in]{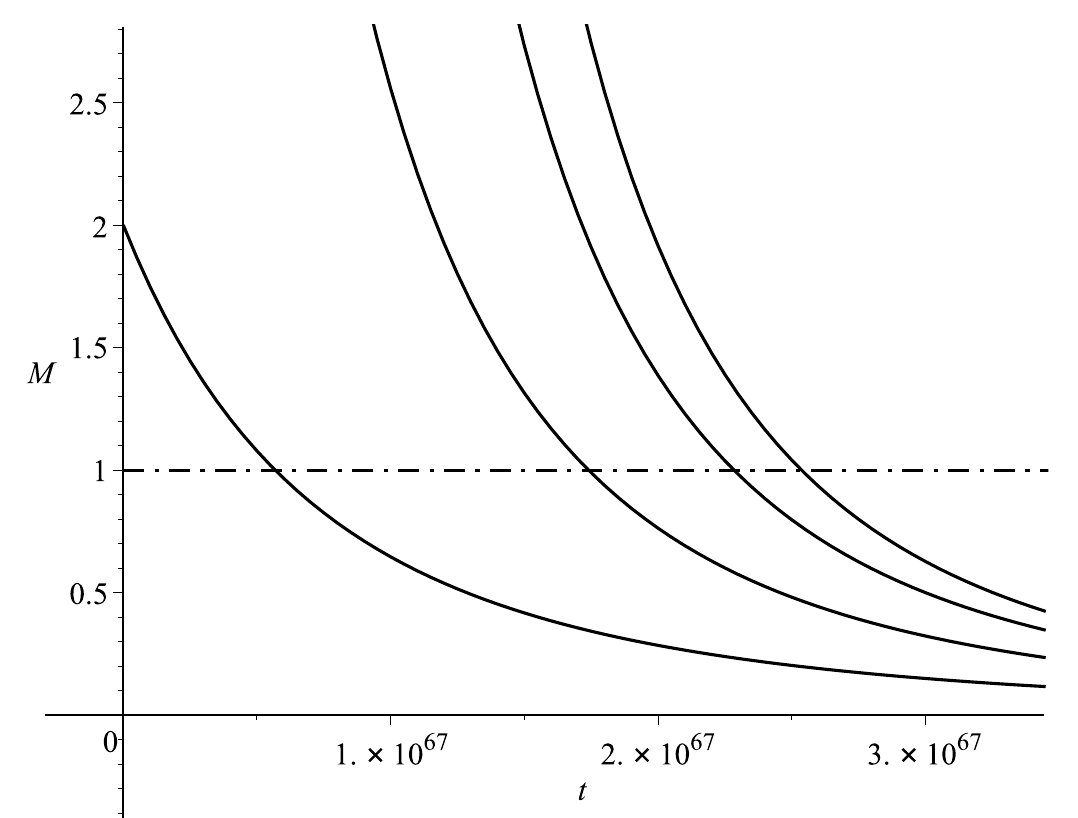} }}
\caption{\textbf{Left:} The evolution of some AdS toral black holes with compactification parameter $K=1$. In this example we set the numerical values of $L$ to unity, and the various initial masses are $20, 200, 2000$ and $20000$, respectively. Although these black holes have seemingly infinite lifetime, they all evaporate down to $M=L$ in a time of order $L^3/\hbar$. Note that the evolution plotted here, which is assuming the geometric optics approximation, should not be trusted around and beyond that point; they were only meant to provide a qualitative picture. 
\textbf{Right:} A closer look of the evolution of these black holes near $M=L$ (dash-dot line).
\label{flat}}
\end{figure}

Let us start from a toral black hole of initial mass $M_0$ and solve for the time $t_*$ it takes to reach a mass $M_*$.
From Eq.(\ref{ODE}), we obtained
\begin{equation}
t_* = \frac{3\mathcal{C} L^{\frac{10}{3}}M_*^{-\frac{1}{3}}}{\hbar} \left[1-\left(\frac{M_*}{M_0}\right)^{\frac{1}{3}}\right].
\end{equation}
We note that the leading term is independent of $M_0$.
This equation holds for any final mass $M_*$, but if $M_* = L$, then it reduces to
\begin{equation}
t_* = \frac{3\mathcal{C}L^3}{\hbar}\left[1-\left(\frac{L}{M_0}\right)^{\frac{1}{3}}\right].
\end{equation}
The second term is negligibly small if $M_0 \gg L$. For a numerical example, see Fig.(\ref{flat}).

If one repeats this exercise in $d=n+2$ dimensions, then if we set $M_* = L^{n-1}$, we would obtain 
\begin{equation}
t_* =  \frac{\mathcal{C}_d (n+1) L^{n+1}}{\hbar}\left[1-\left(\frac{L^{n-1}}{M_0}\right)^{\frac{1}{n+1}}\right],
\end{equation}
where we have emphasized that $\mathcal{C}=\mathcal{C}_d$ is dimensional-dependent.
Note, as a consistency check, that in $d=n+2$ dimensions, mass has dimension of $(\text{length})^{n-1}$, and since the higher dimensional Planck length is defined by 
\begin{equation}
\ell_{\text{Pl}}^n = \frac{G\hbar}{c^3},
\end{equation}
where $G$ is the $d$-dimensional Newton's constant, in the unit such that $G=c=1$ the Planck constant has dimension $(\text{length})^{n}$. So indeed $t_*$ has the dimension of length.\newline \newline
We can also ask a related question:
\begin{itemize}
\item
Given two toral black holes with initial masses $M_0$ and $\widetilde{M}_0$ respectively, how long does it take for $M(t)$ and $\widetilde{M}(t)$ to be as close as $\varepsilon$ apart, for any given $\varepsilon > 0$?  
\end{itemize}
We have, again from Eq.(\ref{sol}), 
\begin{equation}
\big|M(t)^{\frac{1}{3}}-\widetilde{M}(t)^{\frac{1}{3}}\big| = \frac{9\big|\widetilde{M}_0^{-\frac{1}{3}}-{M}_0^{-\frac{1}{3}}\big|}{(\mathcal{B}t + 3M_0^{-\frac{1}{3}})(\mathcal{B}t + 3\widetilde{M}_0^{-\frac{1}{3}})}.
\end{equation}
If $\big|M(t)^{\frac{1}{3}}-\widetilde{M}(t)^{\frac{1}{3}}\big| < \varepsilon $,
then 
\begin{equation}
9\big|\widetilde{M}_0^{-\frac{1}{3}}-{M}_0^{-\frac{1}{3}}\big| < \varepsilon \left[(\mathcal{B}t + 3M_0^{-\frac{1}{3}})(\mathcal{B}t + 3\widetilde{M}_0^{-\frac{1}{3}})\right].
\end{equation}
This yields a quadratic inequality in $t$, which can be solved to yield
\begin{equation}\label{aaa}
t > \frac{1}{2\mathcal{B}^2\varepsilon}\left[-3\varepsilon \mathcal{B} \left(M_0^{-\frac{1}{3}}+ \widetilde{M}_0^{-\frac{1}{3}}\right)\left(1-\sqrt{1-\frac{4}{\varepsilon}\frac{\left(M_0^{-\frac{1}{3}}\widetilde{M}_0^{-\frac{1}{3}}\varepsilon - \big|\widetilde{M}_0^{-\frac{1}{3}}-M_0^{-\frac{1}{3}}\big|\right)}{\left(M_0^{-\frac{1}{3}}+\widetilde{M}_0^{-\frac{1}{3}}\right)^2}}\right)\right].
\end{equation}
It is useful to define a dimensionless quantity $\epsilon$ by $\epsilon:=\varepsilon L^{-\frac{1}{3}}$.

Now, let us consider the special case in which one of these holes have initial mass $\widetilde{M}_0$  being very large, say $(L/\widetilde{M}_0)^{1/3} \sim \epsilon^n $, where $n>1$. 
If we ignore terms $\epsilon^n$ with $n > 1$,
then, the inequality \ref{aaa} reduces to
\begin{equation}
t \gtrsim \frac{3M_0^{-\frac{1}{3}}}{2\mathcal{B}}\left[-1+\frac{2M_0^{\frac{1}{6}}}{\sqrt{\varepsilon}} + \frac{\sqrt{\varepsilon}}{4} M_0^{-\frac{1}{6}}\right].
\end{equation}
For small $\epsilon$, the dominant terms are the first two, and we can re-write the inequality as
\begin{equation}
t \gtrsim  \frac{3\mathcal{C}L^3}{\hbar} \left[\frac{1}{\sqrt{\epsilon}}\left(\frac{L}{M_0}\right)^{\frac{1}{6}}-\frac{1}{2}\left(\frac{L}{M_0}\right)^{\frac{1}{3}}\right].
\end{equation}
This is natural since for the two black holes to get close to $\epsilon$-distance within each other, they need a longer time.
The second term can be ignored if $L \gg M_0$. Indeed if $(L/M_0)^{\frac{1}{3}} \sim \epsilon$ --- and recall that we assume $(L/\widetilde{M}_0)^{\frac{1}{3}} \sim \epsilon^n$, $n>1$ --- then 
\begin{equation}
t \gtrsim \frac{3\mathcal{C}L^3}{\hbar} [1 + O(\epsilon)].
\end{equation}

In addition to toral black holes with compact horizons, one could also study the non-compact, \emph{planar} case,
by taking both $M$ and $K$ to infinity in such a way that the ratio $\mathcal{M}:=M/(4\pi^2 K^2)$ --- the mass density parameter --- remains finite. 
(The mass density is $\mathcal{M}/r_h^2$.)
Then the differential equation \ref{ODE}, 
\begin{equation}
\frac{dM}{dt} \propto -K^{-\frac{2}{3}} M^{\frac{4}{3}}
\end{equation}
reduces to 
\begin{equation} \label{noK}
\frac{d\mathcal{M}}{dt} \propto - \mathcal{M}^{\frac{4}{3}}.
\end{equation}
The same analysis is therefore also applicable to planar black holes, which are \emph{truly infinite} in extent. Their mass densities would still decrease to the order $L$ in a time of order $L^3/\hbar$. 
Indeed, the mass density is the important physical quantity from the point of view of holography, not the mass per se (see, e.g., \cite{kovtun}). 
Note that the explicit $K$-dependence has dropped out if one works with the quantity $\mathcal{M}$, even for the toral case.
A crucial difference between the toral case and the planar case is that: if the periodicity of the torus is comparable to or shorter than the thermal wavelength of the Hawking radiation, then one would expect that there might be some changes in the time scales due to the discreteness of the modes\footnote{The author thanks Don Page for this comment, and the discussions in the next two paragraphs.}. 

To be more specific, note that the $g_{tt}$-component of the metric \ref{toral}, upon substituting in the time-dependent expression of the mass as given by Eq.(\ref{sol}), is, for large initial mass $M_0$, given by
\begin{equation}\label{A}
g_{tt}\approx -\left(\frac{r^2}{L^2}-\frac{\mathcal{A}}{r}\frac{L^{10}}{t^3}\right),
\end{equation}
where $\mathcal{A} \propto \hbar^{-3}$ is a dimensionful constant.  For the simplest case, let us consider sufficiently small $L$ such that $r \gg L$ and the second term of $g_{tt}$ can be neglected, we have $g_{tt} \approx -r^2/L^2$.

The local temperature at fixed value of $r$ is given by the Tolman Law,
\begin{equation}
T_{\text{local}} = \frac{T}{\sqrt{|g_{tt}|}}.
\end{equation}
We have, in this case,
\begin{equation}
T_{\text{local}} \sim \frac{L^3}{rt}.
\end{equation} 
The characteristic wavelength of the radiation is therefore 
\begin{equation}
\lambda \sim rt\frac{\hbar}{L^3}.
\end{equation}
Now, the number of thermal wavelength on a circle (one of the $S^1$-direction of the torus) with periodicity $2\pi K$ is
\begin{equation}\label{discrete}
\frac{2\pi K r}{\lambda} \sim \frac{KL^3}{\hbar t}.
\end{equation}
This is large if $t \ll KL^3/\hbar$, but small if $t \gg KL^3/\hbar$.

Recall that Eq.(\ref{noK}) does not explicitly depend on $K$. In fact, we can work in terms of the re-scaled coordinates
$(\mathfrak{t},\mathfrak{r},\mathfrak{x},\mathfrak{y})=(t/K, Kr, \zeta/K, \xi/K)$, so that $\mathfrak{t}$ has the physical meaning of 
$L$ times the (dimensionless) proper time in the conformally related metric $g[\Bbb{T}^2\text{-AdS}]/r^2$ at infinity with the conformal factor adjusted to give proper spatial circumference $2 \pi$. 
Then, the previous calculation says that the discreteness in the modes becomes important if this re-scaled time $\mathfrak{t} \gg L^3/\hbar$.

Finally, let us remark on the approximation used in Eq.(\ref{A}), in which we have neglected the second term on the right hand side. If we have kept the second term, then Eq.(\ref{discrete}) would have read
\begin{equation}
\frac{2\pi K r}{\lambda} \sim \frac{KL^3}{\hbar t} \left(1-\frac{8\pi \mathcal{A} L^{12}}{r^3t^3}\right)^{-\frac{1}{2}}.
\end{equation}
The factor $\mathcal{J}[r]:=\left(1-{8\pi \mathcal{A} L^{12}}/{r^3t^3}\right)^{-{1}/{2}}$ is of the form $(1-1/x)^{-1/2}$. This is of order unity except for those values of $x$ which are very close to 1. However, in calculating the Tolman temperature, we are interested in some fixed distance sufficiently far away from the black hole horizon. This for two reasons. Firstly, the Tolman temperature diverges at the horizon and the calculation would not make much sense there (see, however, \cite{1508.00312}). Secondly, in the geometric optics approximation, we are interested at the Hawking particles that made it out pass the effective potential of the hole toward the asymptotic observers. Therefore, $x$ indeed should not be too close to unity, and the inclusion of the factor $\mathcal{J}[r]$ only contributes another factor of $O(1)$ to the overall result, and does not affect the qualitative conclusion reached above. Of course, a more detailed analysis is required to establish quantitatively when the discreteness of the Hawking modes become important for a given toral black hole, and how this affects the subsequent evaporation of the hole. This is beyond the scope of the present work. Here, we only point out that in our simple analysis, the discreteness of the mode becomes important at some time governed by, up to some factor, $L^3/\hbar$, which is also around the time the geometric optic approximation breaks down. Therefore, our analysis which does \emph{not} include the discreteness effect into consideration, is nevertheless consistent, in so far as we are only confining our attention to the geometric optics regime.

\section{Black Holes With Negatively Curved Horizons}

Asymptotically locally AdS black holes with negatively curved horizons are quite different from their $k=0$ and $k=+1$ cousins.
Their horizon topologies correspond to the quotients of hyperbolic space $\Bbb{H}^n$ by some discrete group $\Gamma$.
In particular, given a fixed spacetime dimension $d$ and AdS length scale $L$,
these black holes have a \emph{minimum} size $r_{\text{min}}$, which is given explicitly by \cite{9808032, 9705004, 0011097}:
\begin{equation}
r_{\text{min}} = \left(\frac{d-3}{d-1}\right)^{\frac{1}{2}} L,
\end{equation} 
at which point the Hawking temperature vanishes; see Eq.(\ref{temp}). In fact, at this point the mass $M=M_{\text{min}}$ is \emph{negative}, it is:
\begin{equation}
M_{\text{min}} = - \left(\frac{2}{d-1}\right)\left(\frac{d-3}{d-1}\right)^{\frac{d-3}{2}}\frac{L^{d-3}}{\omega_d},
\end{equation}
where
\begin{equation}
\omega_d := \frac{16\pi }{(d-2) V[X^{-1}_{d-2}]}.
\end{equation}

\begin{figure}[!h]
\centering
\includegraphics[width=3.3in]{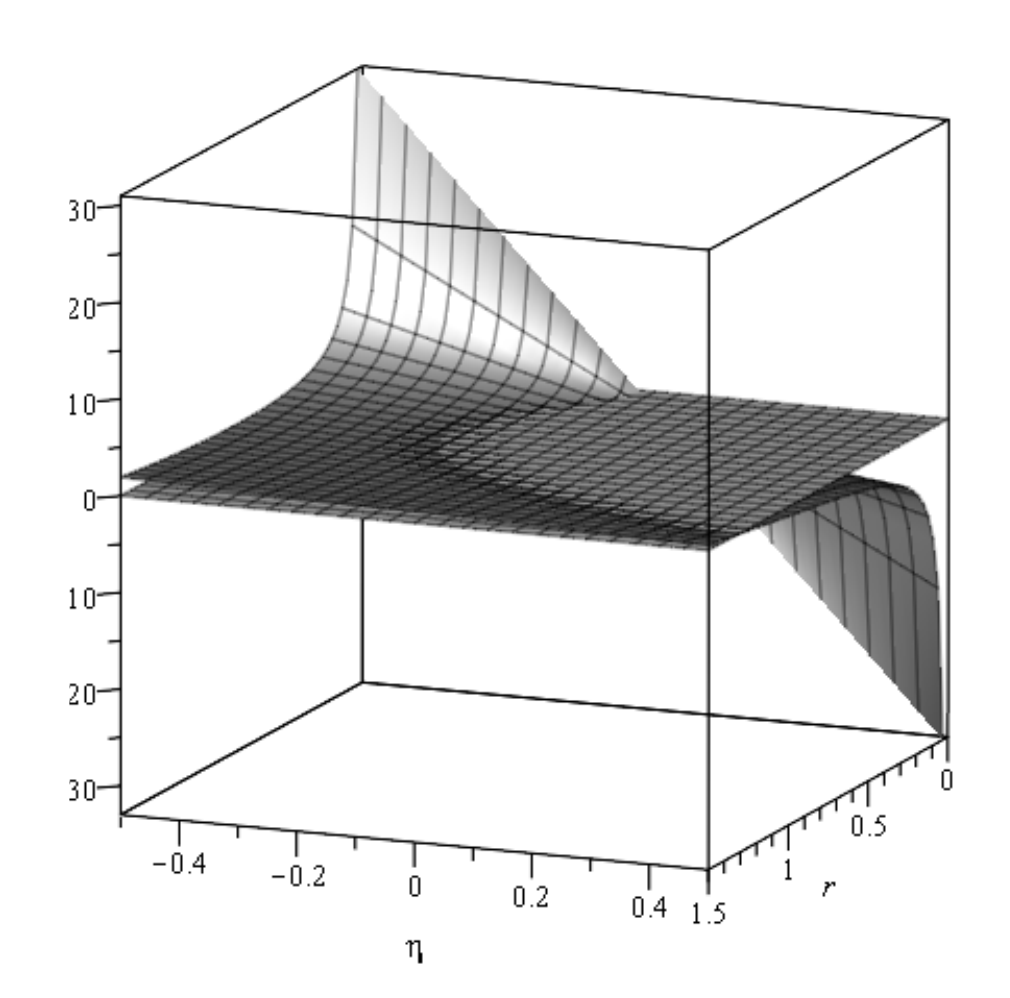}
\caption{The surface defined by $g_{tt}=r^2/L^2 - 1 -2\eta/r$, with $L=1$. The allowed black hole horizons are the positive real roots of  $g_{tt}=0$, shown here as the intersection of the surface with the plane $z=0$, where $z$ labels the vertical axis. Note that for negative mass $\eta$, there are both an outer horizon as well as an inner horizon. \label{horizon}}
\end{figure}

Indeed, in 4-dimensions, the $g_{tt}$ component of the metric tensor \ref{metric} is of the form:
\begin{equation}
\frac{r^2}{L^2}-1-\frac{2\eta}{r}=0,
\end{equation} 
where $2\eta = \omega_d M = 8\pi M/(V[X^{-1}_2])$. 
If $\mathcal{P}:=\eta^2 - L^2/27$ is positive, then 
$g_{tt}$ has a single zero --- and hence only one horizon --- for any given $\eta \geq L/\sqrt{27}$. However, if $\mathcal{P} < 0$, then the allowed values of $\eta$ are in the interval $(-L/\sqrt{27}, +L/\sqrt{27})$. If
$\eta > 0$, there is still only one horizon; but if $\eta < 0$, there are two horizons \cite{9705004}.  An extremal horizon is formed when $M=M_{\text{min}}$. Although these are well-known facts, for completeness we provide an illustration in Fig.(\ref{horizon}).

For a $4$-dimensional black hole with a $2$-dimensional compact orientable event horizon, the Gauss-Bonnet theorem dictates that 
the total amount of the curvature is related to its topology:
\begin{equation}
\int_{S} K~ dA = 2\pi \chi(S)=4\pi (1-g),
\end{equation}
where $K$ is the Gaussian curvature (which is twice the scalar curvature) of the surface $S$, $dA$ the area element, and $g$ its genus. The quantity $\chi(S)=2-2g$ is a topological invariant known as the Euler characteristic.

\begin{figure}[!h]
\centering
\mbox{\subfigure{\includegraphics[width=3.2in]{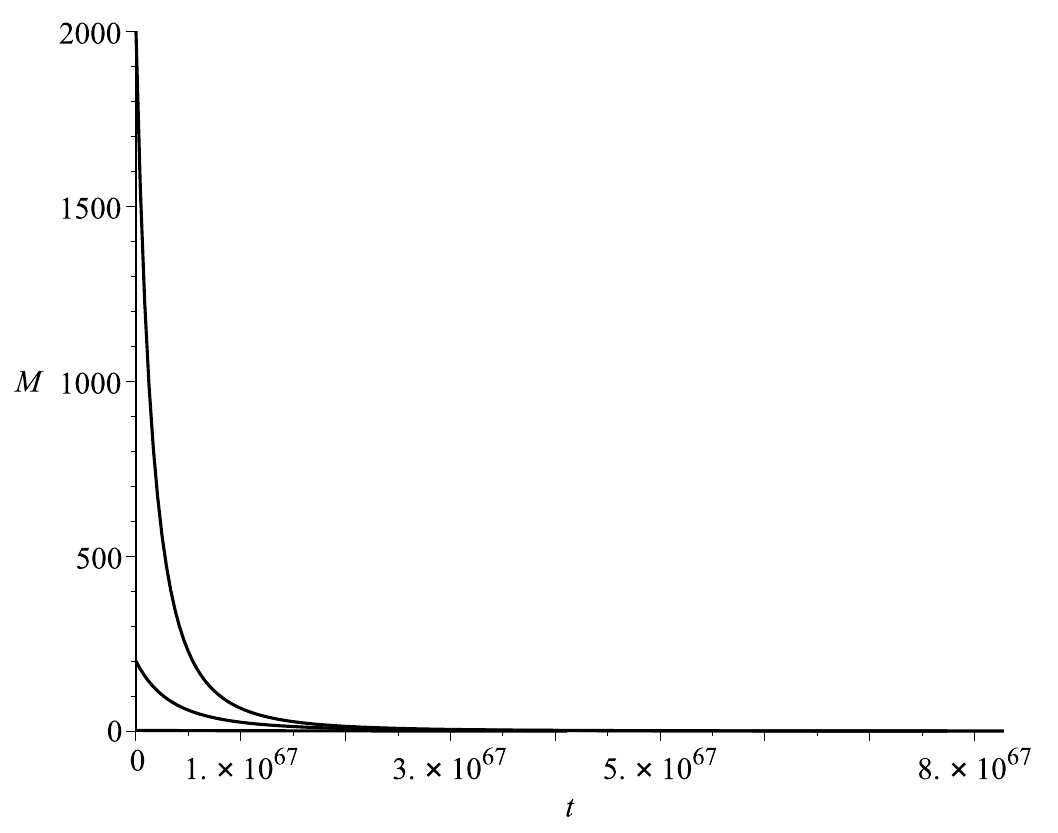}}\quad
\subfigure{\includegraphics[width=3.2in]{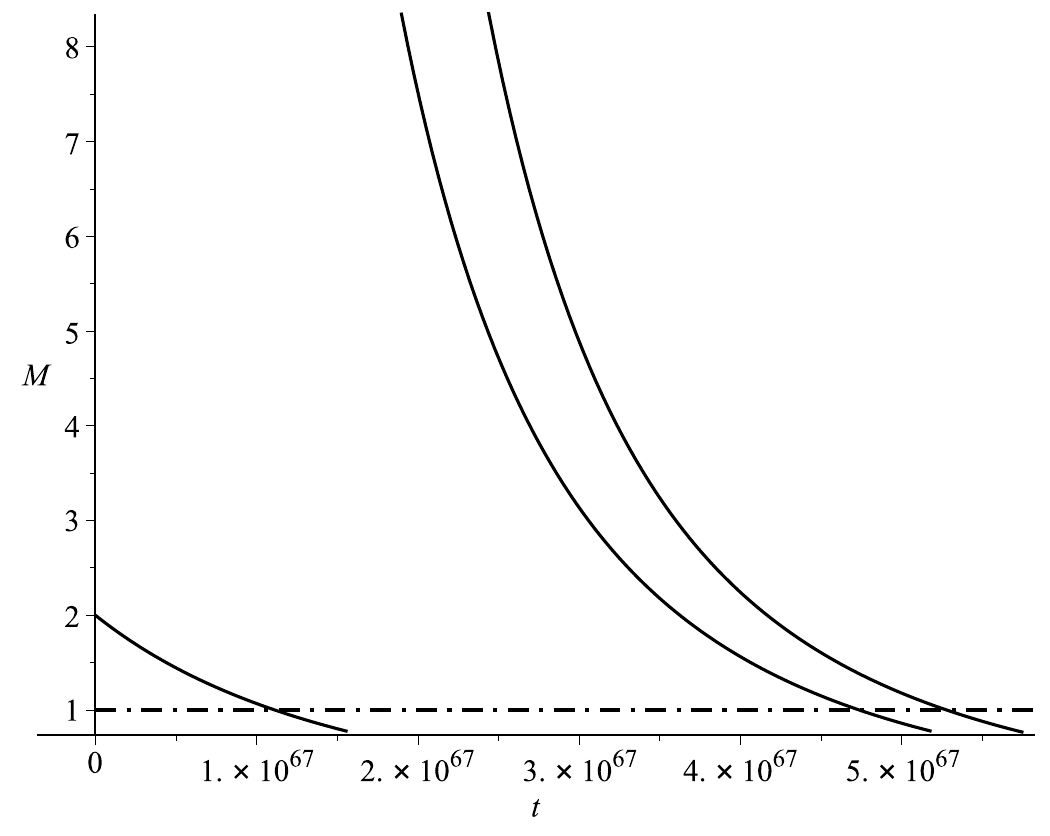} }}
\caption{\textbf{Left:} The evolution of some AdS hyperbolic black holes of genus 2. In this example we set the numerical values of $L$ to unity, and the various initial masses are $2, 200$ and $2000$, respectively. Again, these black holes evaporate down to the size determined by $M=L$ in a time of order $L^3/\hbar$. The evolution should not be trusted quantitatively beyond $M \sim L$, when the geometric optics approximation breaks down.
\textbf{Right:} A closer look of the evolution of these black holes near $M=L$ (dash-dot line).
\label{hyper}}
\end{figure}

Let us consider the case of a compact hyperbolic surface of unit negative curvature with genus 2.
By the Gauss-Bonnet theorem, its dimensionless area is $4\pi$. 
Higher genus implies a larger (dimensionless) area for the emitting surface in the Stefan-Boltzmann law\footnote{In general, the area of the emitting surface also depends on the underlying topology for AdS black holes with \emph{positively} curved event horizons with non-trivial topologies (not only for the flat and negatively curved cases). For example, in 5-dimensions, one could have ``black lenses'' --- black holes with lens space horizon topology $S^3/\Bbb{Z}_p$, $p \in \Bbb{Z}^+$, and so have dimensionless area $2\pi^2/p$. }.
See also \cite{9803061}. 
Much like the toral case, the radiating surface that goes into the Stefan-Boltzmann law still only depends on $L$ and the geometric optic approximation is good for $\eta \gg L$ \cite{9803061}. 
Note that in this particular case, $\eta \equiv M$.
One could then set up a differential equation much like Eq.(\ref{ODE}) to model the evaporation rate. The numerical result, again ignoring the greybody factors, shows that these black holes do evaporate down to $M = L$ in a comparable time scale, given again by $\sim L^3/\hbar$. See Fig.(\ref{hyper}).

\section{Conclusion: Black Hole and AdS Curvature Scale}

In this work, we generalized the recent finding in \cite{1507.02682} to topological black holes in AdS, and found that, at least in general relativity, these black holes share a remarkable property --- arbitrarily large black holes shrink via Hawking radiation to the mass scale set by the cosmological constant, $M=L$, in a finite time of the order $L^3/\hbar$. This is only a qualitative statement, and the correct time scale is probably off by a few order of magnitudes due to the greybody factor, which was ignored in this work. However, even as a qualitative statement, this is a rather remarkable observation, and in some sense, counter-intuitive. 


Since AdS black holes play important roles in the context of holography, it would be interesting to further investigate the implication of this result to the dual field theory. The fact that black holes with an arbitrary mass $M > L$ takes almost the same amount of time to evaporate down to $M=L$ would mean that they take about the same amount of time to reach the critical size at which Hawking-Page phase transition occurs \cite{HP} (which also occurs for toral black holes, to the Horowitz-Myers soliton \cite{9808079,0101134,0108170,0205001}). It is tempting to think that since black holes correspond to some deconfinement phase on the field theory side, this would also mean that confinement-deconfinement transition of the field theory is independent of the mass density. For interesting physical system like quark-gluon plasma, however, the presence of electrical charge means that the current analysis is not directly applicable, and a separate analysis is necessary. See \cite{1403.4886,0910.4456}. It might also be interesting to look into other boundary conditions which would be more useful to model a certain physical system; say, perhaps a partially-reflective boundary condition.

Finally, let us remark that it is very interesting to note that a negative cosmological constant can affect black hole properties in many ways. For an asymptotically flat Schwarzschild black hole, at the classical level there is only one length scale (in the unit $G=c=1$), namely the mass $M$. So it is only natural that this length scale appears in the various properties of the geometry, such as the maximal in-falling time $\pi M$ from the horizon to the spacelike singularity, and the evaporation time scale $M^3/\hbar$. Once there are two length scales $M$ and $L$, it happens that sometimes a combination of $M$ and $L$ characterizes some properties of the black holes, as in the case of the well-known capture cross section of massless particles for AdS-Schwarzschild black hole: $27M^2L^2/(L^2 + 27M^2)$. However, $L$ itself characterizes some important properties of these spacetimes as well, such as the bound for the evaporation time we explored in this work $\sim L^3/\hbar$, and some other physical quantities raised in Section \ref{S1}. 

The lesson here is that physics in AdS can be counter-intuitive, and since general relativity is a \emph{geometric} theory of gravity, we should pay more attention to the effects of the underlying geometry and topology on the various physical properties.


\section*{Acknowledgement}

YCO thanks the Yukawa Institute for Theoretical Physics for hospitality during the molecule-type workshop on ``Black Hole Information Loss Paradox'', YITP-T-15-01; he also thanks Don Page for fruitful discussion, and Brett McInnes for useful comments.


\end{document}